\documentclass[journal]{IEEEtran}
\IEEEoverridecommandlockouts
\usepackage[justification=centering]{caption}
\usepackage{cite}
\usepackage{amsmath,amssymb,amsfonts}
\usepackage{algorithmic}
\usepackage{breakurl}
\usepackage{graphicx}
\usepackage{hyperref}
\usepackage{float}
\usepackage{textcomp}
\usepackage{xcolor}
\begin{document}

\title{SONIC: Sound Optimization for Noise In Crowds}

\author{
    Pranav M N\textsuperscript{\dag}, 
    Gandham Sai Santhosh\textsuperscript{\dag}, 
    Tejas Joshi, 
    S Sriniketh Desikan, 
    Eswar Gupta
    \thanks{\textsuperscript{\dag}Equal contribution.}
    \thanks{All authors are affiliated with the Indian Institute of Technology Madras.\\
    \hspace*{1em}Correspondence to: Gandham Sai Santhosh\\
    \hspace*{2em}(gandhamsanthosh1234@gmail.com)}
}

\maketitle

\begin{abstract}
This paper presents SONIC, an embedded real-time noise suppression system implemented on the ARM Cortex-M7-based STM32H753ZI microcontroller. Using adaptive filtering (LMS), the system improves speech intelligibility in noisy environments. SONIC focuses on a novel approach to noise suppression in audio signals, specifically addressing the limitations of traditional Active Noise Cancellation (ANC) systems. The paper explores various signal processing algorithms in a micro-controller point of view, highlighting various performance factors and which were considered optimal in our embedded system. Additionally we also discussed the system architecture, explaining how the MCU's efficiency was harnessed, along with an in-depth overview of how the audio signals were translated within the processor. The results demonstrate improved speech clarity and practical real-time performance, showing low-power DSP as an alternative to complex AI denoising methods.

\end{abstract}

\begin{IEEEkeywords}
\raggedright
Adaptive filtering, LMS, STM32H753ZI, CMSIS-DSP, Noise suppression, Embedded systems
\end{IEEEkeywords}

\begin{figure}[h]
    \centering
    \includegraphics[width=5cm,height=8.0cm]{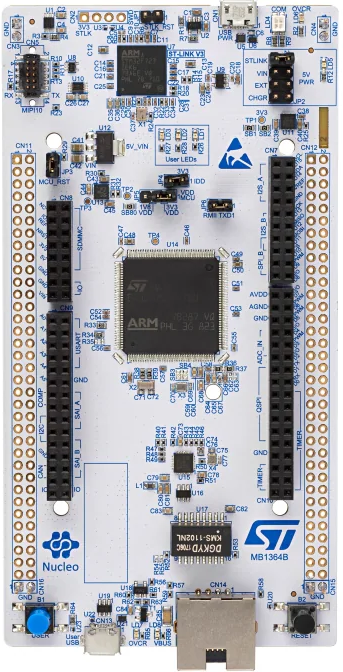}
    \caption{STM32H753ZI (adapted from \cite{stm32h753zi_nucleo})}
    \label{fig:stm32}
\end{figure}

\section{Introduction}
Noise suppression for embedded audio systems is challenging due to power, memory and computational constraints. While ANC and AI-based denoising exist, these either suppress all sounds or require heavy computational resources. This paper presents a real-time DSP-based system using STM32H753ZI MCU with exploration on traditional adaptive filters, meta-heuristic algorithms and beam-forming techniques, achieving noise suppression suitable for low-power devices. Unlike cloud-based or high-power DSP solutions, SONIC runs entirely on-chip, demonstrating that meaningful speech enhancement can be achieved using adaptive filtering algorithms within the strict timing and energy budgets of embedded hardware.\\

Adaptive filtering algorithms, mainly LMS, adjust filter weights to minimize noise, enhancing speech clarity in real-time. A dual-microphone setup was used in this project. This setup follows the traditional adaptive noise filtering framework, where a secondary reference signal, captured via a noise-facing microphone, is used to adaptively filter out unwanted background noise from a primary speech signal. 

This project has involved a thorough investigation of the LMS algorithm and preliminary exploration of the other adaptive filtering techniques like Normalized LMS (NLMS), Recursive Least Squares (RLS), Meta-Heuristic algorithms along with beam-forming techniques also.
\section{Related Work}
Noise suppression on embedded systems has been explored through various methods, primarily traditional adaptive filtering techniques such as LMS and NLMS. These algorithms are popular due to their low computational complexity and ability to adapt in real-time. Prior work demonstrated the feasibility of implementing LMS and NLMS on Cortex-M7-based microcontrollers for educational and experimental purposes. Their work showed promising results in controlled conditions, but primarily focused on didactic platforms rather than real-world robustness.\cite{beccaro2022}

Some studies have also investigated advanced algorithms like RLS, which offer faster convergence but are often too computationally demanding for typical microcontrollers due to their quadratic complexity. As a result, real-time applications on constrained hardware often favor LMS-type algorithms for their simplicity and predictability.\cite{lampl2023}

Meta-heuristic techniques, including Particle Swarm Optimization (PSO) and JAYA, have been proposed for adaptive filtering and speech enhancement. These methods show strong offline performance and can reach high-quality results. However, they are computationally intensive, making them unsuitable for live deployment on low-power embedded devices.\cite{taha2019}

Beamforming is another technique used for spatial filtering using multiple microphones. While effective in certain scenarios, its performance is limited by microphone spacing and array design, particularly on compact hardware platforms. Moreover, adaptive beamforming adds further complexity that is challenging to implement efficiently in real time on embedded systems.\cite{audioxpress2021}

In summary, existing approaches either compromise on performance due to hardware constraints or require computational resources beyond what typical microcontrollers can provide. SONIC aims to bridge this gap by implementing a real-time, low-power noise suppression system that uses adaptive filtering particularly, LMS on a resource constrained embedded platform, achieving practical performance without relying on external processing or cloud-based methods.

\section{Algorithms}
This section chiefly deals with the algorithms that we have reviewed, alongwith the algorithms that have previously been used for the application that we are trying to build. We began our search on Adaptive Filtering algorithms, the ones that were used extensively by the research community. Then, we ventured into other algorithms, primarily experimenting with meta-heuristics, and attempted to establish a comparative rationale between them. Having said that adaptive filtering forms the basis of what we have implemented , a few words on that:

\begin{figure}[h]
    \centering
    \includegraphics[width=0.7\linewidth]{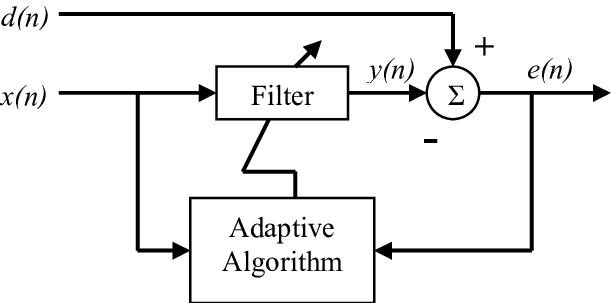}
    \caption{ LMS block diagram }
    \label{fig:adaptivefilter}
\end{figure}

As seen in the figure above, there exist two signals d(n), x(n) and a filter that is changing signal x(n) to signal y(n). d(n) is what we would refer to as the 'desired' signal while x(n) is what we refer as 'reference' signal. To put it clearly, we are envisaging to use 2 mics, one near the source of the speaker and other away from it. Naturally, the noise in both of the inputs would be correlated and it is this correlation that we would utilize.

\begin{figure}
    \centering
    \includegraphics[width=1\linewidth]{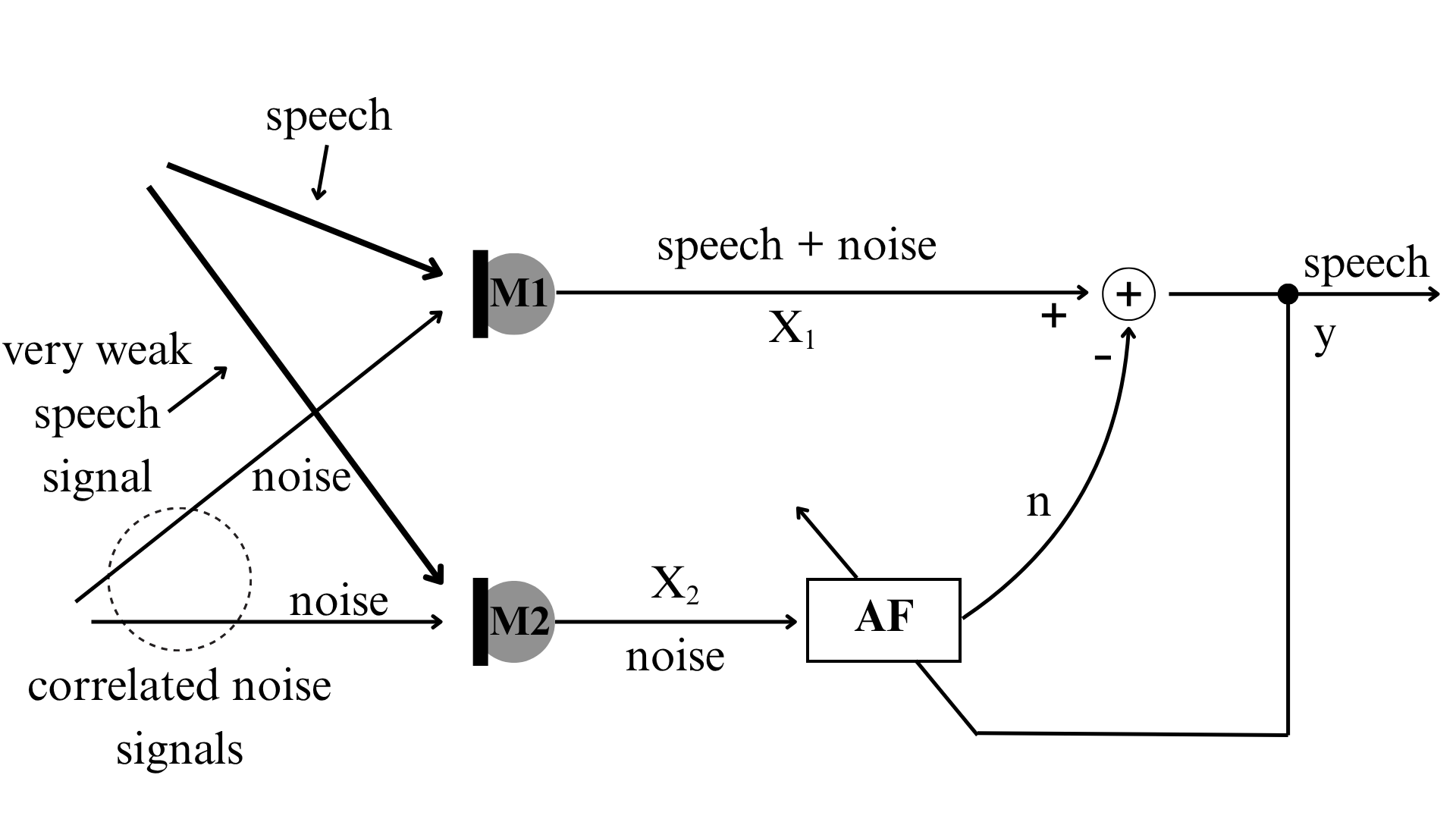}
    \caption{Proposed Microphone Environment}
    \label{fig:micsetup}
\end{figure}

The signal x(n) is transformed by the filter such that, the output y(n) tends towards the noise part of the signal d(n). This essentially is like making a copy of just the noise in signal d(n). The last step would then be to subtract this output from d(n) and obtain the desired speech. The next step in the flow of thought is to understand how the filter works. At its core, the filter is just a bunch of weights that multiply with the delayed versions of the reference signal x(n) and try to bring it closer to the noise in d(n). The weights are updated iteratively and the final output y(n) then converges to the error in O(n) as the number of iterations increase.

\begin{equation}
y[n] = \sum_{k=0}^{L-1} b_k[n] x[n-k]
\end{equation}
\begin{equation}
e[n] = d[n] - y[n]
\end{equation}
\begin{equation}
b_k[n+1] = b_k[n] + \mu e[n] x[n-k]\\
\end{equation}

How the weights update is where the various algorithms play their part. A brief overview of each follows. Note that the first two differ from each other at a more fundamental and a method cumbered with a lot of algebra.

\subsection{Conventional Adaptive Filters}\cite{beccaro2022}

\begin{itemize}
    \item \textbf{LMS:} Simple gradient-based update with O(N) complexity per sample. It converges linearly and can be implemented efficiently in fixed-point or with FPU support. LMS is widely used because it requires minimal computation and memory. Its convergence rate is moderate; the step-size $\mu$ controls the trade-off between speed and stability. In practice on our hardware, CMSIS’s floating-point LMS ran with very low latency (often less than 100 cycles per tap update).
    
    \item \textbf{RLS:} Offers much faster convergence and lower steady-state error than LMS (approaching Kalman filtering performance). However, it requires O(N²) operations (matrix inversions) per step. For a filter of length N, this quickly becomes infeasible on a microcontroller. Simulation shows RLS stabilizes in far fewer iterations, but each iteration is 10–100× more compute-heavy than LMS. Given our real-time constraints and $N \approx 100 taps$, RLS was deemed too computationally expensive despite its superior adaptive performance.
    
    \item \textbf{NLMS: } A normalized LMS variant with dynamic step-size; it offers slightly improved stability over standard LMS at minimal extra cost (one division per block). Since our signals had varying power, NLMS was considered, but its run-time was similar to LMS and gains were marginal for our use case. Refer Fig. 4
\end{itemize}

\begin{figure}
    \centering
    \includegraphics[width=1\linewidth]{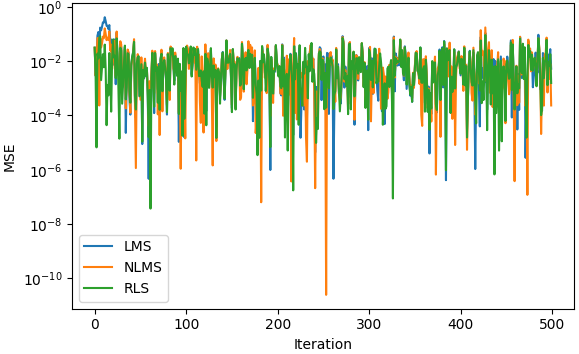}
    \caption{Convergence Comparison}
    \label{fig:convergence}
\end{figure}

\subsection{Meta-Heuristic Algorithms:}
\subsubsection{\textbf{Particle Swarm Optimization (PSO)}:-}

         Let, position if $i^{th}$ particle at time $"t"$ be $x^{i}(t)$ and $v^{i}(t)$ be its corresponding velocity.
         $$x^{i}(t+1)=x^{i}(t)+v^{i}(t+1)$$
         $$v^{i}(t+1)=wv^{i}(t)+c_1r_1(p_{best}^i-x^{i}(t))+c_2r_2(g_{best}-x^i(t))$$

         where,

         $r_1$ and $r_2$ are random numbers and $r_1,r_2 \epsilon[0,1]$

         $w$= weight/inertia/tendency to move

         $c_1$= Cognitive factor $\epsilon[0,1]$ 

         $c_2$= Social factor $\epsilon[0,1]$

         $p_{best}^{i}\rightarrow$ best ever position of $i^{th}$ particle.

         $g_{best}\rightarrow$ best ever positions of all particles. 
         
         This algorithm is a good choice for fast optimization. PSO consistently converged to a low MSE and achieved good SNR improvement\cite{taha2019}. However, its convergence was relatively slower and would be computationally expensive than adaptive filters like LMS or NLMS due to the need to evaluate all particles over multiple iterations. With a significantly higher Execution time PSO was considered unsuitable for real-time processing on embedded hardware\cite{pichardo2022}. While it showed more stable behavior than JAYA, its population-based nature still makes it better suited for offline filter design or pre-training phases rather than live signal processing.\\

\subsubsection{\textbf{JAYA Algorithm:}}

JAYA is a population-based optimization algorithm that requires no algorithm-specific control parameters (such as learning rate or inertia), making it relatively simple to implement and tune\cite{kumar2017}. It updates filter weights by moving each candidate solution toward the best and away from the worst solution in the population, according to:

\[
x'_{j} = x_{j} + r_1(x^{\text{best}}_j - |x_j|) - r_2(x^{\text{worst}}_j - |x_j|)
\]

where \( r_1, r_2 \in [0,1] \) are random numbers, and \( x^{\text{best}}_j \), \( x^{\text{worst}}_j \) are the best and worst performing candidates. This update rule effectively balances exploration and exploitation without the need for explicit hyperparameter tuning. Its inconsistent behavior and high execution time make it unsuitable for real-time embedded DSP. As with other metaheuristics, JAYA is better used in offline filter design rather than adaptive filtering inside an MCU. It is thus better suited for pre-processing stages rather than live noise suppression on resource-constrained hardware.\\

\subsubsection{\textbf{Simulated Annealing}}\cite{ingber1999asa}

Simulated Annealing (SA) is a global optimization heuristic inspired by metal annealing, apt for finding near-optimal filter weights in challenging, non-convex error landscapes. In our project context, the algorithm repeatedly perturbs filter coefficients and occasionally accepts higher-error solutions based on the probability:

\[
P = \exp\left(-\frac{\Delta E}{T}\right)
\]

where \(\Delta E\) is the increase in mean squared error (MSE) and \(T\) is the decaying temperature parameter. Unlike gradient-based LMS/ NLMS and RLS, SA is slower requiring many iterations to converge, which leads to significantly higher execution time (Refer Fig. 6). However, the trade-off is a respectable SNR improvement, often matching RLS or PSO highlighting SA’s strength in escaping local minima and refining global solutions. Within an embedded DSP, SA would face challenges: its sequential nature, heavy use of random sampling, and slow cooling schedule make it inefficient for real-time processing at audio rates of 48 kHz. SA is therefore better suited for offline filter design or tuning phases, rather than being deployed during live audio streaming.

\begin{figure}[h]
    \centering
    \includegraphics[width=0.75\linewidth]{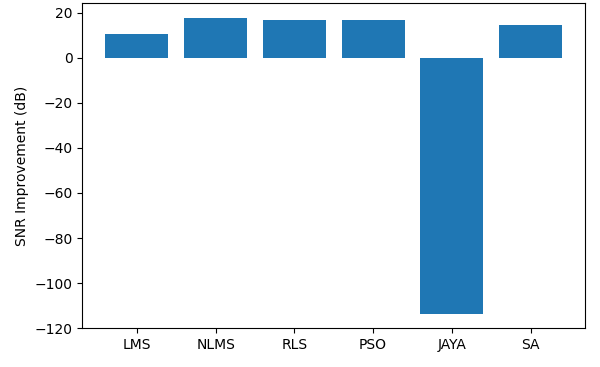}
    \caption{SNR Comparison}
    \label{fig:snrcomparison}
\end{figure}

\subsection{Beam-forming techniques}
In an analog system, the signals are continuous and so is the processing. In a digital system, we have discrete sets of samples. Real time simply means we have to process one set of samples before the next set of samples arrive. This definition of real-time implies that, for a processor operating at a given clock rate, the speed and quantity of the input data determines how much processing can be applied to the data without falling behind the data stream.

For example, if data are sampled at 8 kHz, all processing must be done in $(1/8kHz)$ seconds. All processing has to be done within a finite amount of time.

The signal of each microphone is delayed by a time proportional to the distance between a known target and the microphone, which is done to align the microphone signals. Adding all the signals means that the speech power is enhanced while the noise remains the same.

One of the simplest forms of data-independent beamformers is a delay-and-sum beamformer, where the microphone signals are delayed to compensate for the different path lengths between a target source and the different microphones\cite{audioxpress2021}. This means that when the signals are summed, the target source coming from a certain direction will experience coherent combining, and it is expected that signals arriving from other directions will suffer to some extent from destructive combining\cite{gfai}.

This is not very useful for microphones with small arrays as we need the lengths to be at least a quarter of the wavelengths of the audio involved. There is a second technique called Frost beam formation. The same principle as delay and sum beam-forming but with filter coefficients. This filter used, for example, can be the NLMS filter. Make the multi-channel input a single output audio signal. Let us focus on the case of using two microphones for processing. Using two microphones for the processing means using the fixed geometry of a straight line. This is suitable for focusing on signals from a particular direction\cite{mathworksFrost}. It is found that, typically, the spacing between microphones should be less than or equal to half the wavelength of the highest frequency of interest to avoid spatial aliasing. For speech (100 Hz to 4 kHz), the typical spacing is 5 cm - 20 cm.

Although these Beam-forming techniques can be effective in specific situations, its success heavily depends on the spacing between microphones and the overall array configuration such as factors that pose constraints on compact hardware. Additionally, implementing adaptive beamforming increases computational complexity, making real-time execution on embedded systems difficult. 

\subsection{Conclusion}
These algorithms search for optimal filter weights through iterative randomness. They are non-deterministic and require many evaluations. For example, PSO might use 20–50 particles over 100+ iterations to optimize filter taps – that’s thousands of function evaluations for one output sample. In simulations we saw PSO/JAYA yielding high-quality solutions offline, but even coarse settings needed far more time than real-time’s 1/48,000s per sample. Meta-heuristics were considered but not deployed due to non-deterministic computation and high runtime.” Given these trade-offs, we chose LMS for the embedded filter. LMS provides a predictable fixed cost per sample and the MCU’s FPU/DSP extensions keep it fast. 

\begin{figure}[h]
    \centering
    \includegraphics[width=1.0\linewidth]{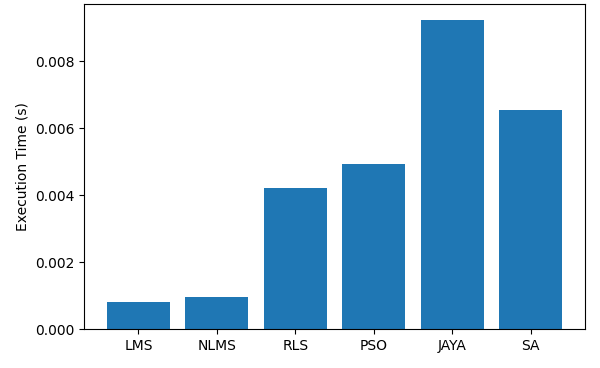}
    \caption{Runtime}
    \label{fig:runtime}
\end{figure}

\section{System Architecture}
\subsection{Our Computational Engine \cite{stm32h753zi_datasheet}} 
The STM32H753ZI was chosen for its unmatched combination of high performance and audio-centric features. It uses a 480 MHz Arm Cortex‑M7 core with a double-precision FPU and full DSP instruction set, yielding 1027 DMIPS. The device has large on-chip memory (2MB flash, 1MB RAM) and multi-layer bus architecture for peak throughput. Crucially, its peripherals include multiple serial audio interfaces: six SPIs (three of which can run in duplex I2S mode) plus four full SAIs, as well as four DFSDM channels for PDM mics. These allow dual I2S microphone inputs and stereo outputs. In hardware, the H7’s four DMA controllers (MDMA and regular DMAs) support circular buffering so audio streams can flow directly between I/O and RAM without CPU overhead. The block diagram in Fig. 7 highlights the M7 core, caches, DMA engines.

\begin{figure}[h]
    \centering
    \includegraphics[width=1.0\linewidth]{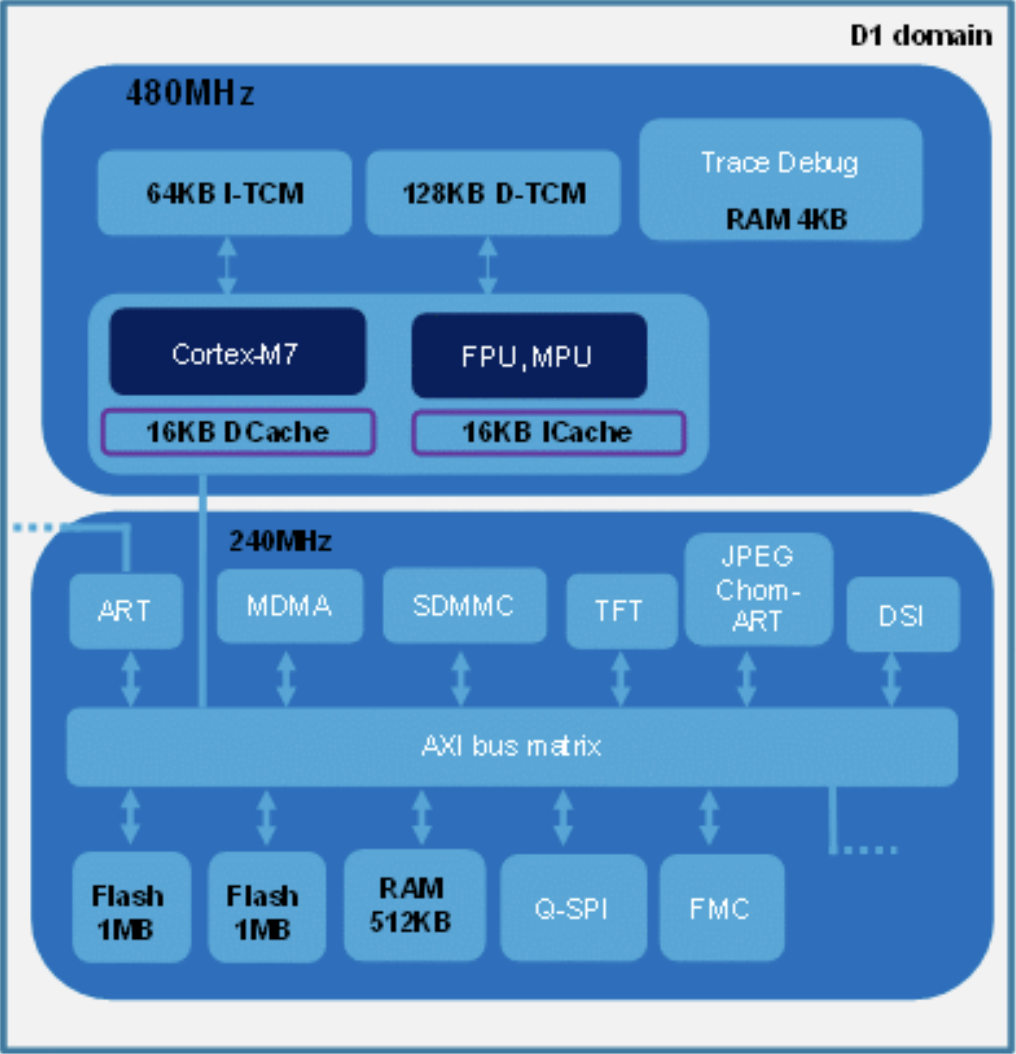}
    \caption{D1 sub system (adapted from \cite{an5557_dualcore_architecture})}
    \label{fig:architecture}
\end{figure}

The Cortex‑M7 core features a DP-FPU and SIMD MAC units, ideal for fast FIR-filtering operations. CMSIS-DSP routines (e.g. \texttt{arm\_lms\_f32()}) utilize this FPU to execute multiply-accumulate operations in hardware. Compared to older Cortex‑M4 MCUs (e.g. STM32F4), the H7 delivers roughly twice the floating-point throughput and includes a larger register file and caches, enabling sustained high-speed signal processing.
The H7’s audio interfaces were vital for SONIC. It supports dedicated I2S blocks and full-duplex SAIs, enabling simultaneous multi-channel audio I/O. We used two I2S input channels (one per microphone) and one output channel in stereo mode\cite{daga2024}.
With a 480 MHz core clock and flexible PLLs, the MCU comfortably meets 48 kHz streaming (requiring a 1–12 MHz I2S master clock) and headroom for computation. For example, PLLI2S can be configured so that 256 Fs (for Fs = 48 kHz) clock the I2S block. Such high clocking plus fast memory allows running the adaptive filter on large blocks without data underflow.

\subsection{Understanding at a Processor Level}
The H7’s FPU critically accelerates the LMS filter. The Cortex‑M7 FPU can execute single-precision MACs in a few cycles (often one per instruction), and has pipeline/SIMD that allow multiple operations per cycle. The CMSIS-DSP \texttt{arm\_lms\_f32()} is hand-optimized to exploit these units\cite{stAN4841}. In contrast, fixed-point Q15/Q31 LMS versions require manual scaling and post-shifting and still risk overflow. On our tests, \texttt{arm\_lms\_f32()} achieved significantly higher throughput than the Q15 version\cite{cmsisDSP}. Internally, the FPU-based code uses instructions like VMLA (vector multiply-accumulate) to process four samples per loop in parallel. The net effect is that the FPU has made 24-bit audio filtering real-time feasible. In sum, the double-precision FPU and SIMD extensions of the M7 chip are the reason real-time 24-bit floating-point filtering runs efficiently on this MCU.\\

The I2S data format and DMA configuration are subtle. On the data bus, a 24-bit audio sample is actually transmitted in a 32-bit frame (with 8 bits of padding). We initially configured the DMA to transfer 32-bit words, but this misaligned the 24-bit samples, resulting in gibberish outputs. Changing the DMA to 16-bit half-word mode fixed it: each 32-bit I2S frame was handled as two 16-bit transfers. In effect, the first half-word carried the lower 16 bits of the 24-bit sample, and the second half-word (ignored or fixed) contained padding, but the data aligned. This choice yielded the correct PCM values (at the cost of discarding the highest 8 bits, effectively using a 16-bit audio resolution). In general, one must match the I2S data-size setting, DMA data width, and buffer layout carefully. Circular DMA mode (with 16-bit transfers) allowed the sample stream to loop endlessly- once DMA reached the end of the buffer, it wrapped back to the start\cite{dmaEmbedded}. The CPU never participates in copying data; it only processes full or half-buffer interrupts, while DMA automatically refills the other half. This is the “circular double-buffer” scheme that ensures real-time overwriting of old samples.\\

We chose a 48 kHz sample rate because it is a standard audio rate (widely supported by microphones and DACs) and simplifies clock generation. Many MEMS microphones natively support 48 kHz, and STM32’s audio PLLs can easily generate the required clocks for 48 kHz (for example, setting PLLI2S to 256×Fs yields an exact MCLK). Using standard rates like 48 kHz avoids fractional dividers or resampling. Clock tree can be used to compute these frequencies: e.g. an 8 MHz crystal → PLLI2S (×24 = 192 MHz) → I2S clock 192/4 = 48 MHz bit clock at 32 bits/sample. In short, 48 kHz was chosen for compatibility and because the MCU’s PLLs can produce it exactly; using a non-standard rate would complicate the PLL settings. Refer Fig. 8.

\begin{figure}[h]
    \centering
    \includegraphics[width=1.0\linewidth]{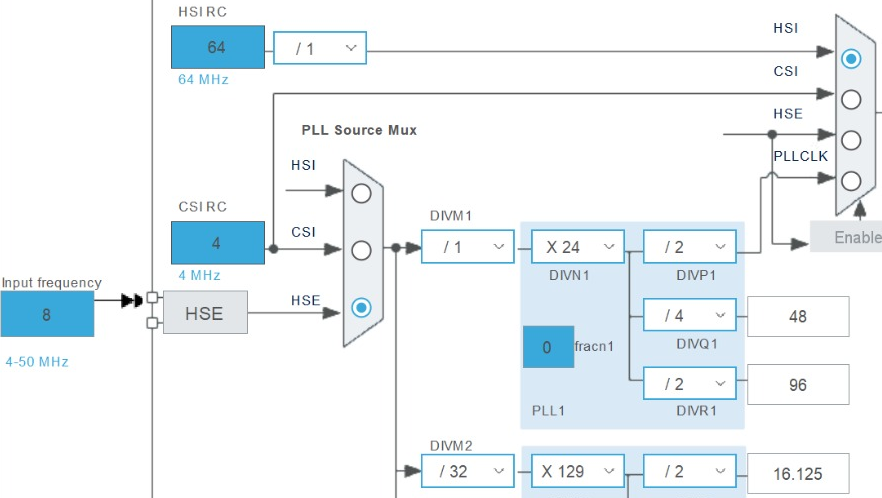}
    \caption{Clock Tree}
    \label{fig:clocktree}
\end{figure}

\section{Hardware Overview}
Two MEMS I2S microphones feed I2S channels of STM32H753ZI; output PCM streams are LMS filtered, then sent via another I2S channel where an external Class-D audio amplifier drives this output to the embedded speaker. 

\begin{figure}[h]
    \centering
    \includegraphics[width=1.0\linewidth]{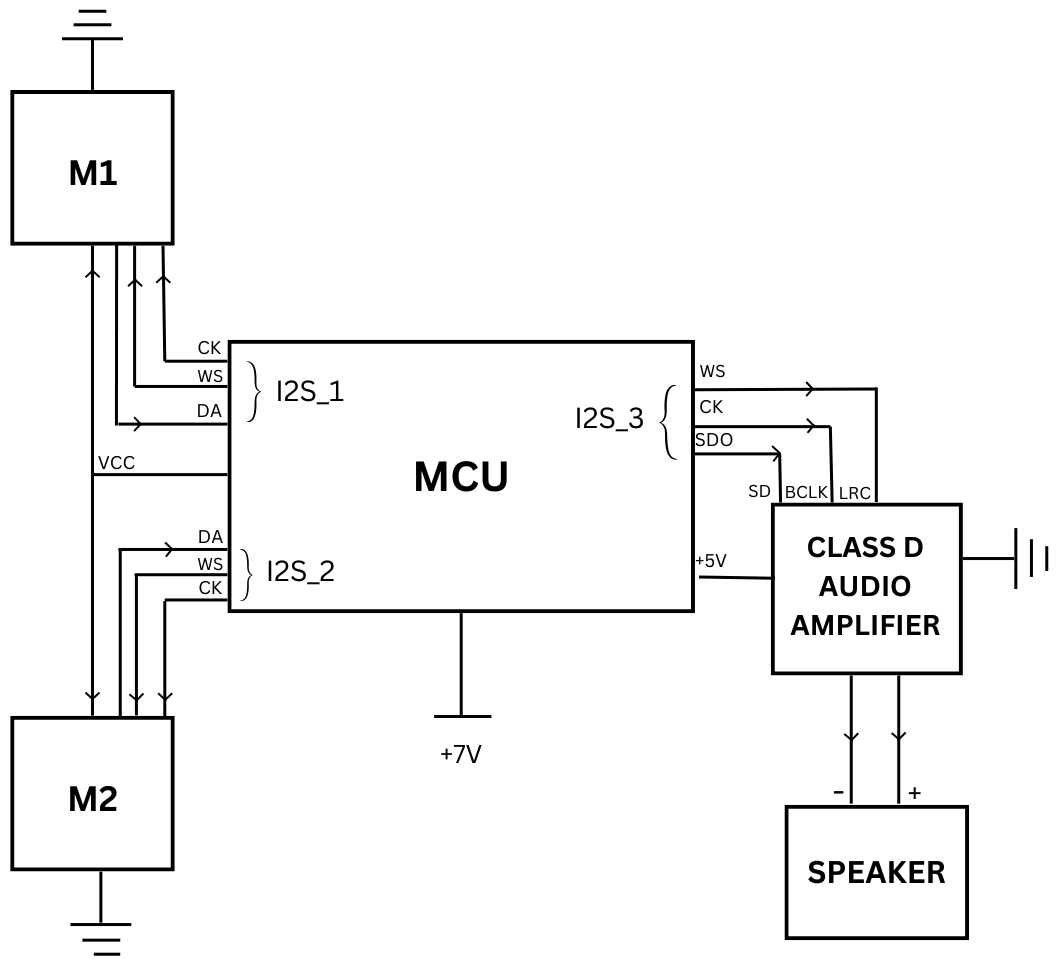}
    \caption{Embedded Layout}
    \label{fig:embeddedlayout}
\end{figure}

We used the MCU’s dual-DMA controllers in circular (ping-pong) mode. Incoming audio from I2S is written into a RAM buffer continuously, while the CPU processes the other half. Likewise, outgoing data to the I2S uses a separate DMA channel in circular mode. This double-buffering ensures no data gaps. In effect, while DMA fills buffer B0 with new mic samples, the CPU filters the previous buffer B1, and vice versa, keeping both the input and output streams uninterrupted. Refer Fig. 9.

\begin{figure}[h]
    \centering
    \includegraphics[width=1.0\linewidth]{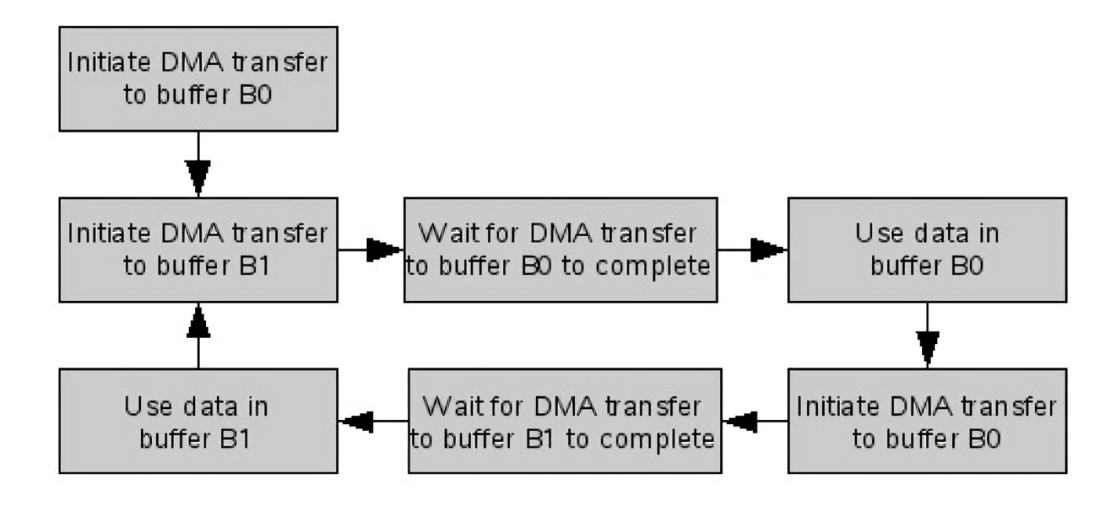}
    \caption{Double Buffering (adapted from \cite{ieee_4404939})}
    \label{fig:doublebuffering}
\end{figure}

Two I2S channels were used in parallel: one per microphone. The STM32 supports multiple I2S/SAI peripherals, so each MEMS mic’s PCM data could be received on its own data line. The STM32’s I2S modules were clocked by a dedicated audio PLL (e.g. PLLI2S), providing the master clock (MCLK) and bit-clock (BCLK) for precise 48 kHz sampling. DMA requests on the I2S periphs transfer each audio sample into memory automatically. Our board has two 12‑bit DAC channels. We found the 12-bit DAC insufficient for 24‑bit PCM input: mapping 24-bit data into 12-bit output caused severe amplitude truncation\cite{stm32f4filter}. Indeed, the team observed “DAC resolution is a hardware constraint limiting output quality without external amplification”. Adding an off-chip Class‑D audio amplifier restored full dynamic range and drive strength.

\section{Understanding Audio Flow}

Before real-time DSP, we implemented the filter using stored .WAV files on an SD card to understand how these audio signals streamed. We used the STM32’s high-speed SDMMC interface instead of SPI mode because SDMMC supports 4- or 8-bit data and clock up to 48–125 MHz, giving much higher throughput. Our tests confirmed SDMMC could stream audio data rapidly enough to the MCU RAM to simulate real-time input. We used the FatFS library for filesystem access. Initially we experienced “mount failures” with FatFS, which was later traced to the need for hardware pull-ups: the STM32’s internal 30–50 k$\Omega$ pull-ups on the SD lines were too weak. Adding external 10 k$\Omega$ pull-up resistors on CMD and DAT lines fixed the issue. In summary, SDMMC was chosen for speed (vs. slower SPI/1-bit SDIO) and proper pull-ups are required for reliable card detection and data transfer. Once the data was streaming from SDMMC into RAM, the same LMS code (ARM CMSIS-DSP) was applied to each audio buffer exactly as in real-time mode (just with CPU-driven reads instead of DMA). This testing verified algorithm functionality before moving to live I2S input.\\ 

In software, WAV files were read in chunks (blocks) of PCM samples, which were then fed to the LMS filter. In our implementation, we used a filter length of 96 taps with a block size of 256 samples, in line with ARM’s CMSIS-DSP guidelines. Accordingly, the state buffer size was set to numTaps + blockSize – 1 = 351. We selected a step-size parameter of $\mu = 0.01$, which provided a good trade-off between convergence speed and stability for speech-band signals. With this configuration, the \texttt{arm\_lms\_f32()} function adapted effectively within a few hundred samples and exhibited no signs of divergence during real-time operation. The signal flow is: each new block of reference (noise) samples x[n] is stored by DMA in RAM, then \texttt{arm\_lms\_f32()} is called to compute the output y[n] and error e[n] = d[n]–y[n], updating coefficients concurrently.

\begin{figure}[h]
    \centering
    \includegraphics[width=1.0\linewidth]{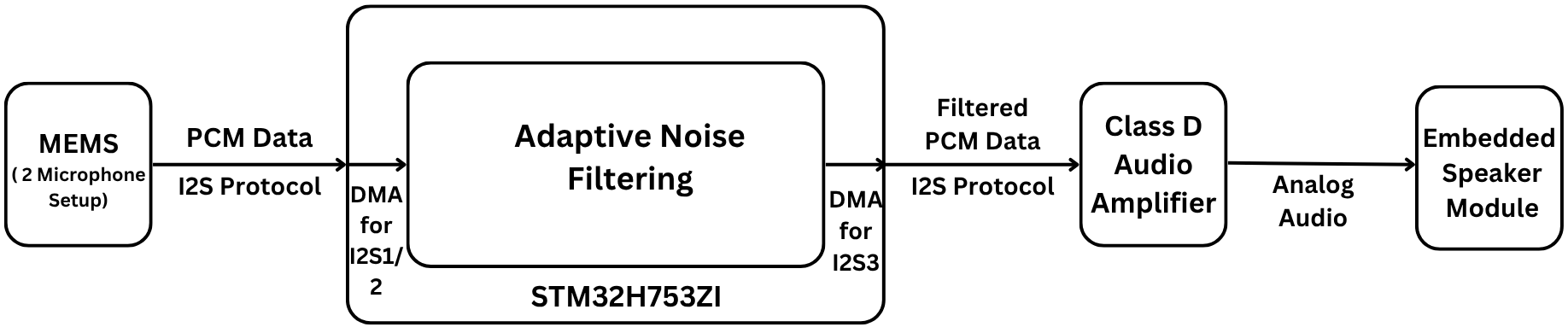}
    \caption{Data Flow}
    \label{fig:dataflow}
\end{figure}

We selected MEMS I2S microphones (which present already-decoded PCM data) to simplify the signal chain. Some digital MEMS mics output a raw pulse-density-modulated (PDM) bitstream requiring on-chip decimation filters (e.g. ST’s DFSDM blocks) to produce PCM. While the H7 does include DFSDM filters for PDM decoding, using a mic with native PCM output removes this extra processing stage. In other words, our chosen microphone included its own sigma delta modulator and decimator to deliver 24-bit PCM words over I2S, so the MCU simply treated it as a standard I2S ADC. This avoids any need for the PDM→PCM conversion library. From an industrial perspective, using a PCM-encoded mic reduces software complexity and latency.\\

I2S (Inter-IC Sound) is a synchronous serial protocol specifically for audio data. Each I2S bus consists of a word-select (WS), bit-clock (BCLK), and data line, and optionally a master clock (MCLK). We configured three I2S channels in total: one left/right pair for the microphones and one for output. The STM32’s I2S/SAI blocks support DMA transfers directly. We set up DMA in circular mode so that as soon as one buffer fill completes an interrupt, DMA automatically continues into the next buffer half. This means the CPU never has to intervene for each sample – it only processes a block when a half-buffer is full. In effect, double-buffered DMA ensures continuous audio: while DMA writes new samples into half the buffer, the CPU reads and filters the other half. We explicitly used three I2S peripheral instances: two in master-receive mode (for the two mics) and one in master-transmit. Each was clocked by a precise audio PLL to achieve the 48kHz rate.

\begin{figure}[h]
    \centering
    \includegraphics[width=1.0\linewidth]{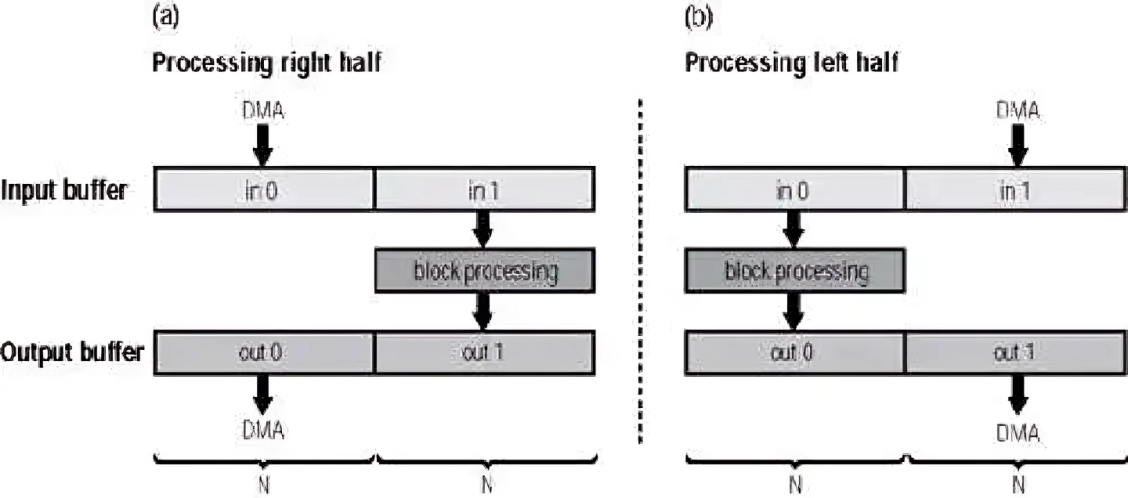}
    \caption{Circular DMA (adapted from \cite{dmaEmbedded})}
    \label{fig:circulardma}
\end{figure}

Initially, using the on-chip 12-bit DAC introduced severe quantization. Since our input was 24-bit audio, the top 8 bits were effectively lost, causing noise and truncation. This was confirmed by experiments and noted in the Discussion: “Hardware constraints (e.g., DAC resolution) limit output quality without external amplification”. The solution was to route the filtered signal through an external Class‑D amplifier (which accepts 16-bit DAC input at full scale) and drive a speaker. The Class‑D’s higher resolution and power greatly improved audio quality. Regarding buffering: we used double-buffering on the MCU. This means the DMA and CPU share a buffer that is split in half. When one half is being filled by DMA, the CPU processes the other half. In STM32 HAL terms, we started DMA with MultiBuffer mode, and each half-buffer triggers a “half-transfer complete” interrupt. Thus, at any time one buffer half holds the newest data and the other is being processed. This decouples the DMA from the CPU. In practice, our code had callbacks for “Buffer 0 ready” and “Buffer 1 ready,” and the LMS filter ran on whichever half was just filled, while the I2S DMA output used the opposite half. This architecture guarantees that new audio always overwrites the oldest data (real-time streaming) without glitches. 

\section{Results and Evaluation}
\subsection{Objective/ Subjective Metrics}
SNR, STOI, and PESQ were measured pre- and post-filtering on noisy speech samples\cite{picovoice2023}. After processing, both the unfiltered (noisy) and filtered signals generated by our real-time LMS system were stored directly onto the SD card via the STM32H753ZI board itself. These recordings were then evaluated using actual outputs from our embedded setup (not simulated) to calculate objective metrics like SNR, STOI, and PESQ through dedicated analysis scripts. A clean, pre-recorded speech sample served as the reference for these comparisons. Importantly, the improvements we observed such as a 7 dB gain in SNR, a STOI increase from 0.69 to 0.84, and a PESQ rise from 1.69 to 2.55 reflect the real-world performance of our hardware implementation. These results demonstrate the system’s ability to perform effective noise suppression in practical conditions, independent of any simulated environment.

\begin{table}[H]
\caption{Performance Metrics}
\centering
\scalebox{1.2}{ 
\begin{tabular}{|c|c|c|}
\hline
\textbf{Metric} & \textbf{Before} & \textbf{After} \\
\hline
SNR (dB) & +5 & +12 \\
STOI & 0.69 & 0.84 \\
PESQ & 1.69 & 2.55 \\
MOS & - & 4.47 \\
\hline
\end{tabular}
}
\end{table}

For subjective evaluation, we conducted a live real-time demo during a research event at our university, where 122 participants spoke into the microphone and listened to the processed output through the speaker. Each rated the filtered audio using the standard MOS scale (1 to 5). The final MOS score was 4.47 out of 5, indicating clear improvement in speech quality and noise reduction. This real-time feedback reflected the system’s practical effectiveness.

\section{Conclusion}
We presented SONIC, a real-time noise suppression system built on the STM32H753ZI using LMS adaptive filtering. Our work shows that even with limited compute and memory budgets, it is possible to achieve effective speech enhancement on embedded platforms without the need for AI or cloud-based processing.

The significance of SONIC lies in its feasibility on low-power MCUs and its deterministic, low-latency behavior, making it suitable for battery-powered and real-time applications like smart speakers, hearing aids, or IoT voice interfaces. While meta-heuristic algorithms like PSO and JAYA offered better offline performance, their high execution time and non-determinism rendered them unsuitable for real-time deployment on MCUs.

Looking ahead, a promising direction is to translate this architecture into a custom ASIC design, allowing tighter integration and further reduction in power and latency ideal for use inside consumer-grade sound devices. Such a hardware-software co-design could make embedded noise suppression mainstream, offering an alternative to high-cost or opaque AI models in scenarios that demand transparency, efficiency, and edge autonomy.

\bibliographystyle{IEEEtran}

\begin{thebibliography}{99}

\bibitem{stm32h753zi_nucleo}
STMicroelectronics, ``STM32H753ZI Nucleo-144 Board,'' [Online]. Available: \url{https://www.st.com/en/evaluation-tools/nucleo-h753zi.html}. [Accessed: June 15, 2025].

\bibitem{beccaro2022}
M. Beccaro, A. Zambolin, and D. Ravazzolo, "Low-cost didactic platform for real-time adaptive filtering: Application on noise cancellation," \textit{IEEE Access}, vol. 10, pp. 19579–19588, 2022.

\bibitem{lampl2023}
G. Lampl, A. Sedaghat, and G. Gatti, "Noise cancellation with LMS, NLMS and RLS filtering algorithms to improve the fault detection of an industrial measurement system," \textit{Measurement}, vol. 212, 2023.

\bibitem{taha2019}
B. Taha and A. M. S. Rahman, "Speech Enhancement Based on Adaptive Noise Cancellation and Particle Swarm Optimization," \textit{Indonesian Journal of Electrical Engineering and Computer Science}, vol. 15, no. 2, pp. 930–937, 2019.

\bibitem{pichardo2022}
J. A. Pichardo, R. V. Carrillo, and E. Osornio-Rios, "A Novel PSO-Based Adaptive Filter Structure with Switching Selection Criteria for Active Noise Control," \textit{Sensors}, vol. 22, no. 2, pp. 569–577, 2022.

\bibitem{kumar2017}
S. Kumar and S. Singh, "Jaya-FLANN based adaptive filter for mixed noise suppression from ultrasound images," \textit{International Journal of Imaging Systems and Technology}, vol. 27, no. 4, pp. 1031–1038, 2017.

\bibitem{ingber1999asa}
L. Ingber, ``Adaptive simulated annealing for optimization in signal processing,'' in \emph{Proc. IEEE Congr. Evol. Comput.}, 1999, pp. --.

\bibitem{picovoice2023}
Picovoice, "Measuring Speech Quality: An Overview of Metrics for Voice Enhancement," \textit{Picovoice Blog}, 2023. [Online]. Available: \url{https://picovoice.ai/blog/speech-quality/}

\bibitem{stm32h753zi_datasheet}
STMicroelectronics, \textit{STM32H753ZI: Arm\textsuperscript{\textregistered} Cortex\textsuperscript{\textregistered}-M7 32-bit MCU datasheet}, 2023. [Online]. Available: \url{https://www.st.com/resource/en/datasheet/stm32h753zi.pdf}

\bibitem{stAN4841}
STMicroelectronics, "Digital signal processing for STM32 microcontrollers using CMSIS," \textit{Application Note AN4841}, 2018. [Online]. Available: \href{https://www.st.com/resource/en/application_note/an4841-digital-signal-processing-for-stm32-microcontrollers-using-cmsis-stmicroelectronics.pdf}{STMicroelectronics AN4841}

\bibitem{an5557_dualcore_architecture}
STMicroelectronics, \textit{STM32H745/755 and STM32H747/757 Lines Dual-Core Architecture}, Application Note AN5557, 2022. [Online]. Available: \href{https://www.st.com/resource/en/application_note/an5557-stm32h745755-and-stm32h747757-lines-dualcore-architecture-stmicroelectronics.pdf}{STMicroelectronics AN5557}

\bibitem{cmsisDSP}
Arm Ltd., "CMSIS-DSP: LMS Filters," \textit{CMSIS Documentation}, 2022. [Online]. Available: \url{https://arm-software.github.io/CMSIS_5/DSP/html/group__LMS.html}

\bibitem{gfai}
gfai tech GmbH, "Delay-and-Sum Beamforming in the Time Domain," \textit{Knowledge Base}, 2021  [Online]. Available: \url{https://www.gfaitech.com/knowledge/faq/delay-and-sum-beamforming-in-the-time-domain}

\bibitem{audioxpress2021}
Michael Tuttle and Jakob Vennerød, "Microphone Array Beamforming with Optical MEMS Microphones" \textit{audioXpress Magazine}, 2024.  [Online]. Available: \url{https://audioxpress.com/article/microphone-array-beamforming-with-optical-mems-microphones}

\bibitem{ieee_4404939}
S. Haykin, "Adaptive Filter Theory," in \textit{IEEE Transactions on Acoustics, Speech, and Signal Processing}, vol. 30, no. 1, pp. 140–142, Feb. 1982. [Online]. Available: \url{https://ieeexplore.ieee.org/stamp/stamp.jsp?tp=&arnumber=4404939}


\bibitem{mathworksFrost}
MathWorks, "Frost Beamformer Block," \textit{MathWorks Documentation}, 2022. [Online]. Available: \url{https://www.mathworks.com/help/phased/ref/frostbeamformer.html}

\bibitem{daga2024}
Anonymous, "LMS-Based Real-Time Embedded Noise Canceller using STM32H7 MCU," in \textit{Proc. DAGA 2024}, pp. 1–4. [Online]. Available: \url{https://pub.dega-akustik.de/DAGA_2024/files/upload/paper/88.pdf}

\bibitem{stm32f4filter}
R. Rinaldi, "Analysis Of The Effect Filter Order Number On Noise Canceller System Using STM32F4," \textit{ResearchGate}, 2023. [Online]. Available: \url{https://www.researchgate.net/publication/366793531_Analysis_Of_The_Effect_Filter_Order_Number_On_Noise_Canceller_System_Using_STM32F4}

\bibitem{dmaEmbedded}
J. Faludi, "Using Direct Memory Access Effectively in Media-Based Embedded Applications – Part 3," \textit{Embedded.com}, 2021. [Online]. Available: \href{https://www.embedded.com/using-direct-memory-access-effectively-in-media-based-embedded-applications-part-3/}{embedded.com}

\end{thebibliography}

\end{document}